\documentclass[a4paper,11pt]{article}
\usepackage{pos}
\usepackage{braket}
\usepackage{bbold}
\usepackage{subcaption}
\usepackage{graphicx} 
\graphicspath{{figs/}}

\title{Hadronic Parity Violation from Twisted Mass Lattice QCD }

\author*[a]{Nikolas Schlage}
\author[a]{Marcus Petschlies}
\author[a]{Aniket Sen}
\author[a]{Carsten Urbach}

\affiliation[a]{Helmholtz-Institut für Strahlen- und Kernphysik, Rheinische Friedrich-Wilhelms-Universität Bonn, Nussallee 14-16, 53115 Bonn, Germany}

\emailAdd{schlage@hiskp.uni-bonn.de}

\abstract{We present results for an exploratory lattice calculation of
  the leading parity-violating pion-nucleon coupling $h_\pi^1$. Based
  on the PCAC relation we use a parity-conserving Lagrangian and focus
  on the techniques to determine the nucleon matrix elements of the
  effective 4-quark operators. For our study we employ an ensemble of
  twisted mass fermions with $260$ MeV pion mass. Barring mixing with
  lower-dimensional operators and renormalization at this stage, we
  discuss our estimate for $h_\pi^1$.}

\FullConference{%
	The 39th International Symposium on Lattice Field Theory (Lattice2022),\\
	8-13 August, 2022 \\
	Bonn, Germany 
}

\begin{document}
\maketitle

\section{Introduction}

Even in the present time hadronic parity violating ($PV$) processes in the low energy regime are poorly understood. 
On the theoretical side, hadronic $PV$ processes are described in the model provided by Desplanques, Donoghue, and Holstein (DDH) in Ref.~\cite{Desplanques:1979hn}  
simply by 1-meson exchanges. Restricting further to the $\Delta I =
1$ sector, 
long-range single pion contribution reduces the set of seven immanent phenomenological 
couplings to a single one, the pion-nucleon coupling $h^1_{\pi}$.

As described in~\cite{Kaiser:1989ah,Meissner:1998pu,deVries:2015pza}, further efforts have been made to determine more precisely the coupling constants, including the pion-nucleon coupling, in the framework of chiral effective field theory.

While the first experimental results for the pion-nucleon coupling constant have high uncertainties and are even partially compatible with zero, 
see e.g. Ref.~\cite{Page:1987ak}, a breakthrough was achieved in 2018 by the
NPDGamma collaboration~\cite{NPDGamma:2018vhh}, where the DDH coupling was estimated
as $h_\pi^1 = ( 2.6 \pm 1.2\,(\mathrm{stat.}) \pm 0.2\,(\mathrm{sys.})) \times 10^{-7}$.

This sparked renewed interest in the theory determination of the DDH coupling. Its evaluation requires
as non-perturbative input nucleon-to-nucleon-pion matrix elements of $\Delta I = 1$ 4-quark operators from electro-weak neutral current products.
In a pioneering effort, a first such lattice determination was performed by Wasem already in 2012~\cite{Wasem:2011tp}. 
However, as pointed out in Refs.~\cite{Walker-Loud,Guo:2018aiq}, using a single lattice spacing and volume, heavier-than-physical pion mass $M_{\pi} \approx 390 $ MeV, neglect of quark loop diagrams, and lack of renormalization severely hamper a rigorous comparison to experiment.

Another strategy to determine the DDH coupling $h_\pi^1$ was recently
put forward in Ref.~\cite{Feng:2017iqb}, based on a soft pion theorem.
Here, we report on our investigation of this approach based on nucleon matrix elements of parity-conserving 4-quark operators.
We calculate for the first time the bare matrix elements of all the
relevant 4-quark operators for one ensemble of $N_f=2+1+1$ twisted
mass fermions~\cite{Frezzotti:2000nk} at maximal
twist~\cite{Frezzotti:2003ni} at pion mass $M_{\pi} = 260$ MeV
generated by the Extended Twisted Mass Collaboration
(ETMC)~\cite{ExtendedTwistedMass:2020tvp,Alexandrou:2018egz}.

%

\section{Nucleon Matrix Elements with PCAC Techniques}

%
$h_{\pi}^1$ is defined by the effective electro-weak parity-odd interaction Lagrangian $\mathcal{L}_{PV}^w$ as
\begin{equation}
h_\pi^1 = -\frac{i}{2 m_N}\, \lim_{p_\pi \to 0} \braket{n\, \pi^+ | \mathcal{L}_{PV}^w | p}.
\label{eq:h_pi_1_1}
\end{equation}
Using the partially-conserved axial current (PCAC) relation as suggested in Ref.~\cite{Feng:2017iqb} and restricting to leading order perturbation theory the soft-pion matrix elements are approximable by
\begin{equation}
\lim_{p_\pi \to 0} \braket{n\, \pi^+ \,| \, \mathcal{L}_{PV}^w \,|\, p} \approx
-\frac{\sqrt{2} i}{F_\pi} \, \braket{p|\mathcal{L}_{PC}^w\,|\,p} = \frac{\sqrt{2} i}{F_\pi}\, \braket{n|\mathcal{L}_{PC}^w\,|n}\,,
\end{equation}
where the effective parity-conserving ($PC$) Lagrangian $\mathcal{L}_{PC}^{w}$ is constructed from the operators $\theta_k^\prime$ analogously to its $PV$ counterpart, i.e.
\begin{equation}
\mathcal{L}_{PC}^w = -\frac{G_F}{\sqrt{2}}\, \frac{\sin^2 (\theta_W)}{3} \left( \sum_{k=1}^3 C_k^{(1)} \, \theta_k^{(\ell)^\prime} + \sum_{k=1}^4 S_k^{(1)} \, \theta_k^{(s)^\prime} \right).
\label{eq:L_PC_w}
\end{equation}
Here, $C_k^{(1)}$ and $S_k^{(1)}$ are the next-to-leading order Wilson coefficients at the scale $1~\text{GeV}$~\cite{Tiburzi:2012hx}, $(\ell)$ labels the $PC$ operators built entirely from light $u$ and $d$ quark fields and $(s)$ labels those $PC$ operators which additionally contain strange quarks. Furthermore, $G_F$ is the Fermi constant and $\theta_W$ the weak mixing angle.
According to Ref.~\cite{Feng:2017iqb}, in leading-order chiral perturbation theory the pion-nucleon coupling can be estimated as
\begin{equation}
h_\pi^1 \approx -\frac{(\delta m_N)_{4q}}{\sqrt{2}\, F_\pi}.
\label{eq:h_pi1}
\end{equation}
Hence, given the induced neutron-proton mass shift
\begin{equation}
(\delta m_N)_{4q} = (m_n - m_p)_{4q} = 
\frac{\Braket{p|\mathcal{L}_{PC}^w(0)|p}}{m_N},
\label{eq:delta_mN}
\end{equation}
the pion-nucleon coupling can be calculated on the lattice from plain,
parity even nucleon matrix elements.

\section{Quark Flow Diagrams}

For the calculation of nucleon matrix elements Eq.~(\ref{eq:delta_mN}),
different quark flow diagrams are required, each representing nucleons
at initial and final time $t_i$ and $t_f$, respectively, and a quark insertion at intermediate time $t_c$.
The proton and neutron at source and sink time are interpolated by the 3-quark operators
\begin{equation*}
p = \epsilon_{abc}\, \left[ u_a^T\, C\gamma_5\, d_b \right]\, u_c 
\qquad \text{~and~} \qquad n = \epsilon_{abc} \, \left[ d_a^T\, C\gamma_5 \, u_b \right]\, d_c 
\end{equation*}
with charge conjugation matrix $C$.
In our work, $PC$ 4-quark operators $\theta_k^{(\ell/s)\prime}$ are
inserted in between. We express all of them in terms of
quark-bilinear products.

\begin{figure}[htbp!]
\centering
\begin{minipage}[t]{0.31\textwidth}
\includegraphics[width=\textwidth]{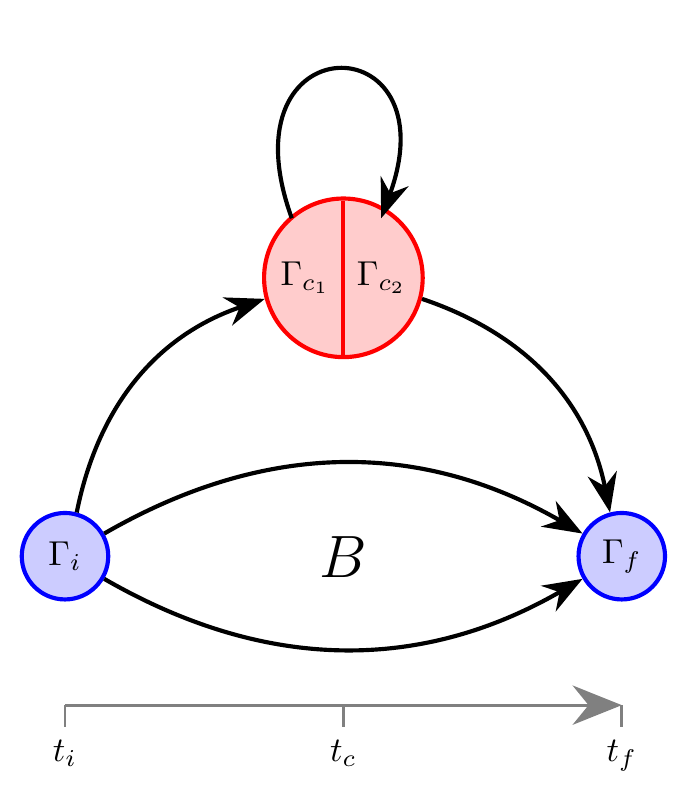}
\end{minipage}
\begin{minipage}[t]{0.31\textwidth}
\includegraphics[width=\textwidth]{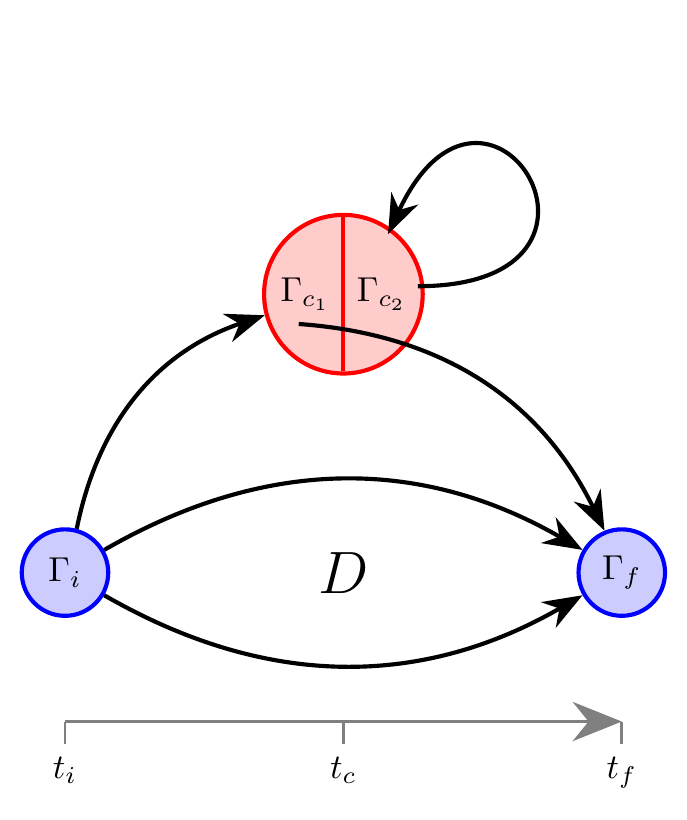}
\end{minipage}
\begin{minipage}[t]{0.31\textwidth}
\includegraphics[width=\textwidth]{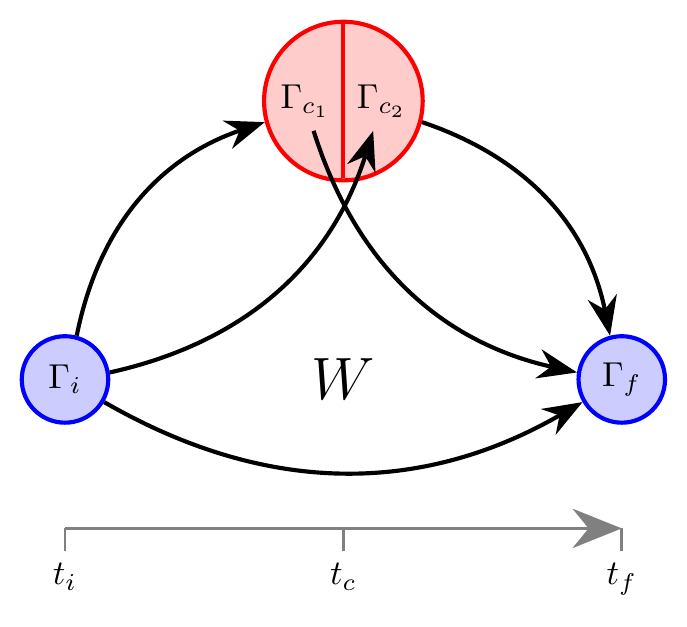}
\end{minipage}\caption{From left to right: Quark flow diagrams of type $B$, $D$ and $W$}
\label{fig:quark_flow_diagrams}
\end{figure}

The 4-quark operators at the insertion point $x_c = (t_c, \vec{x}_c)$
read in general
\begin{equation*}
	\bar{q}(t_c, \vec{x}_c)\, \Gamma_{c_1}\, q(t_c, \vec{x}_c)\, \bar{q}(t_c, \vec{x}_c)\, \Gamma_{c_2}\, q(t_c, \vec{x}_c).
\end{equation*}
More specifically, we use the seven operators~\cite{Feng:2017iqb}
\begin{equation}
\begin{gathered}
\theta_1^{(\ell)\prime} = \bar{q}_a\, \gamma_\mu \, \mathbb{1} \, q_a\, \bar{q}_b\, \gamma^\mu \, \tau^3 \, q_b,  \quad \theta_2^{(\ell)\prime} = \bar{q}_a\, \gamma_\mu \, \mathbb{1} \, q_b \, \bar{q}_b\, \gamma^\mu \, \tau^3\, q_a,\\
\theta_3^{(\ell)\prime} = \bar{q}_a \, \gamma_\mu \, \gamma_5 \, \mathbb{1} \, q_a \, \bar{q}_b \, \gamma^\mu \, \gamma_5 \, \tau^3 \, q_b,
\end{gathered}
\label{eq:theta_ell}
\end{equation}
and
\begin{equation}
\begin{gathered}
\theta_1^{(s)\prime} = \bar{s}_a \, \gamma_\mu \, s_a \, \bar{q}_b \, \gamma^\mu \, \tau^3 \, q_b, \quad \theta_2^{(s)\prime} = \bar{s}_a \, \gamma_\mu \, s_b \, \bar{q}_b \, \gamma^\mu \, \tau^3 \, q_a,\\
\theta_3^{(s)\prime} = \bar{s}_a \, \gamma_\mu \, \gamma_5 \, s_a \, \bar{q}_b \, \gamma^\mu \, \gamma_5 \, \tau^3\, q_b, \quad \theta_4^{(s)\prime} = \bar{s}_a \, \gamma_\mu \, \gamma_5 \, s_b \, \bar{q}_b \, \gamma^\mu \, \gamma_5 \, \tau^3 \, q_a,
\end{gathered}
\label{eq:theta_s}
\end{equation}
where $a$ and $b$ are color indices. Inserting the light interpolators Eq.~(\ref{eq:theta_ell}) makes the 3-point correlator decompose into three different diagram types called $B$, $D$ and $W$ shown in Fig.~\ref{fig:quark_flow_diagrams}.
On the other hand, in the strange quark sector only diagrams of type $B$ and $D$ are relevant.
Unlike the $W$-type diagram, which is loop-free, the $B$- and $D$-type diagrams each have a loop attached to the $x_c$ insertion point. The difference between the latter two is that in the $B$-type diagram the loop extends from $\Gamma_{c_1}$ to $\Gamma_{c_2}$, while in the $D$-type diagram it is connected only to $\Gamma_{c_2}$, cf. Fig.~\ref{fig:quark_flow_diagrams}.

The two major challenges in this calculation are (1) the signal-to-noise ratio for the diagram types $B$ and $D$ containing a quark-loop;
and (2) the quark-loop induced mixing on the lattice from those same diagrams with operators of lower dimension. 
Our report focusses on (1), whereas the latter must be addressed by proper non-perturbative renormalization and is on-going work.

\section{Matrix Elements with Feynman-Hellmann-Theorem}

To determine the nucleon matrix elements we apply the Feynman-Hellmann-Theorem (FHT) as described e.g. in Ref.~\cite{Bouchard:2016heu}
with perturbation $\mathcal{L}_{PC}^w$ inducing
the shift of the (pure QCD) spectrum. In particular the nucleon mass is obtained from the positive-parity projected 2-point function

%
\begin{equation}
  C^\text{2pt} (t) = \mathrm{Tr}\left[ \frac{1}{2}\,\left(\mathbb{1} + \gamma_0   \right)\,\braket{N(t)\,\bar N(0)} \right]
  \overset{t \mathrm{~large} }{\longrightarrow}
\frac{1}{2 m_N}\, \braket{\Omega \,|\, N | n } \, \braket{n \, | \, \bar{N} | \Omega}\, e^{-m_N  t} 
\quad + \mathrm{~excited~states}
\label{eq:C2pt_t_to_infty}
\end{equation}
via the effective mass
%
\begin{equation}
m_\text{eff} (t, \tau) = \frac{1}{\tau} \mathrm{arccosh} \left( \frac{C^\text{2pt}(t+\tau) + C^\text{2pt}(t-\tau)}{2 C^\text{2pt}(t)} \right).
\label{eq:m_eff}
\end{equation}
%
Following the FHT method with perturbed lattice action
\begin{equation}
S = S_\text{LQCD} + \lambda \sum_x \mathcal{L}_{PC}^w (x)
\label{eq:action-fht}
\end{equation}
%
%
%
gives
\begin{equation}
  \left. \frac{\partial m_{\mathrm{eff}}}{\partial \lambda}\right|_{\lambda = 0} \overset{t \to \infty}{=} -\frac{\Braket{N|\mathcal{L}_{PC}^w(0)|N}}{2 m_N}.
  \label{eq:fht}
\end{equation}
%
Likewise, the latter matrix elements can be determined using FHT ratios.
Specifically, taking the partial derivative of Eq.~(\ref{eq:m_eff}) with respect to $\lambda$ and using the $\lambda$-dependence of the action
in Eq.~(\ref{eq:action-fht}), the FHT ratio is given by
\begin{equation}
R_{k,X}^{(j)\prime}(t, \tau) = \frac{1}{\tau} \frac{z^{(j)}}{\sqrt{\left(z^{(j)}\right)^2 - 1}} \left[ \frac{C_{k, X}^{\text{3pt}(j)}(t+\tau) + C_{k, X}^{\text{3pt}(j)}(t-\tau)}{C^{\text{2pt}}(t+\tau) + C^{\text{2pt}}(t-\tau)} - \frac{C_{k, X}^{\text{3pt}(j)}(t)}{C^{\text{2pt}}(t)} \right],
\label{eq:FHT_ratio}
\end{equation}
where we defined
%
$z = \left( C^{\text{2pt}}(t+\tau) + C^{\text{2pt}}(t-\tau) \right) / \left( 2 C^{\text{2pt}}(t) \right)$ and
\begin{align}
  C_{k, X}^{\text{3pt}(j)}(t) &=  \mathrm{Tr}\left[ \frac{1}{2}\,\left(\mathbb{1} + \gamma_0   \right)\,
    \braket{N(t)\,\,\sum\limits_{x_c} \,\theta^{(j)\prime}_{k,X}(x_c) \,\, \bar N(0)}
\right]\,.
  \label{eq:fht-3pt}
\end{align}
Here, $j = \ell,\, s$ indicates whether the light or strange quark sector is considered, $k$ specifies the 4-quark operator at insertion and $X = B,\, D,\, W$ the diagram type.
Furthermore, the 3-point function corresponding to the partial derivative of the 2-point function with respect to $\lambda$ is denoted with $C^\text{3pt}$.
In the limit of large $t$ the target matrix elements can be estimated from the FHT ratio by
\begin{equation}
R_{k,X}^{(j)\prime}(t,\tau) \overset{t \to \infty}{=} \frac{\braket{N|\theta_{k,X}^{(j)\prime}|N}}{2 m_N} \qquad \forall\, \tau \ge 1.
\label{eq:FHT_ratio_2}
\end{equation}
It should be emphasized that for $t \to \infty$ the $\tau$ dependence vanishes.
In the following, for the purpose of comparing with Ref.~\cite{Wasem:2011tp}, we perform an analogous approximate determination of the coupling as in said reference, by using
only the $W$-type diagram and staying at the level of bare matrix elements. Based on Eq.~(\ref{eq:h_pi1}) we denote this estimate by
%
\begin{equation}
h_\pi^1(W, \text{bare}) \approx \frac{G_F\, (\hbar c / a)^2 \, \sin^2(\theta_W)}{3\, a F_\pi} \sum_{k=1}^3 C_k^{(1)}\, \mathcal{M}_{k,W}^{(\ell)\prime} \qquad \forall\, \tau \ge 1 \,.
\label{eq:h_pi1_only_W}
\end{equation}
Here, $\mathcal{M}_{k,W}^{(\ell)\prime}$ is an estimate of the nucleon matrix element as given in Eq.~(\ref{eq:FHT_ratio_2}) obtained from a constant fit to the plateau region in the corresponding ratio data.

\section{Lattice Simulation}

All lattice calculations are performed on the basis of the $N_f =
2+1+1$ gauge field ensemble cA211.30.32 provided by the
ETMC~\cite{ExtendedTwistedMass:2020tvp}. The corresponding lattice 
action combines the Iwasaki improved gauge action on the one hand and
fermion action terms for light and heavy quark doublets on the
other~\cite{Alexandrou:2018egz}. The details of the ensemble are: $L^3 \times
T=32^3\times64$, $a=0.097\ \mathrm{fm}$, $M_\pi = 261\ \mathrm{MeV}$ and
$M_\pi L = 4$, $m_N = 1028(4)\ \mathrm{MeV}$.
In total, we use 1262 gauge configurations.
All quark flow diagrams are composed of both point-to-all propagators and sequential propagators.
A description of the calculation techniques used for the different
quark flow diagrams is provided in Ref.~\cite{Sen:2021dcb}.

For noise reduction it is beneficial to increase the number of stochastic samples. In the case of $W$-type diagrams we use 8 such samples. Since these calculations are numerically very expensive for the loop diagrams, we use only a single stochastic sample when calculating $B$- and $D$-type diagrams. However, in order to reduce the noise nevertheless, we use 8 different source coordinates for each of these diagrams. In the case of the $W$-type diagrams, 2 source coordinates are sufficient.

Source and sink smearing is applied to all propagators, where we use APE smearing for the gauge fields~\cite{APE:1987ehd} and Wuppertal smearing for the fermion fields~\cite{Gusken:1989qx}.

\section{Results}
\subsection{Nucleon Matrix Elements}

\begin{figure}[htbp!]
\centering
\hfill
\subfloat[\bf{(a)}]{\includegraphics[width=0.497\textwidth]{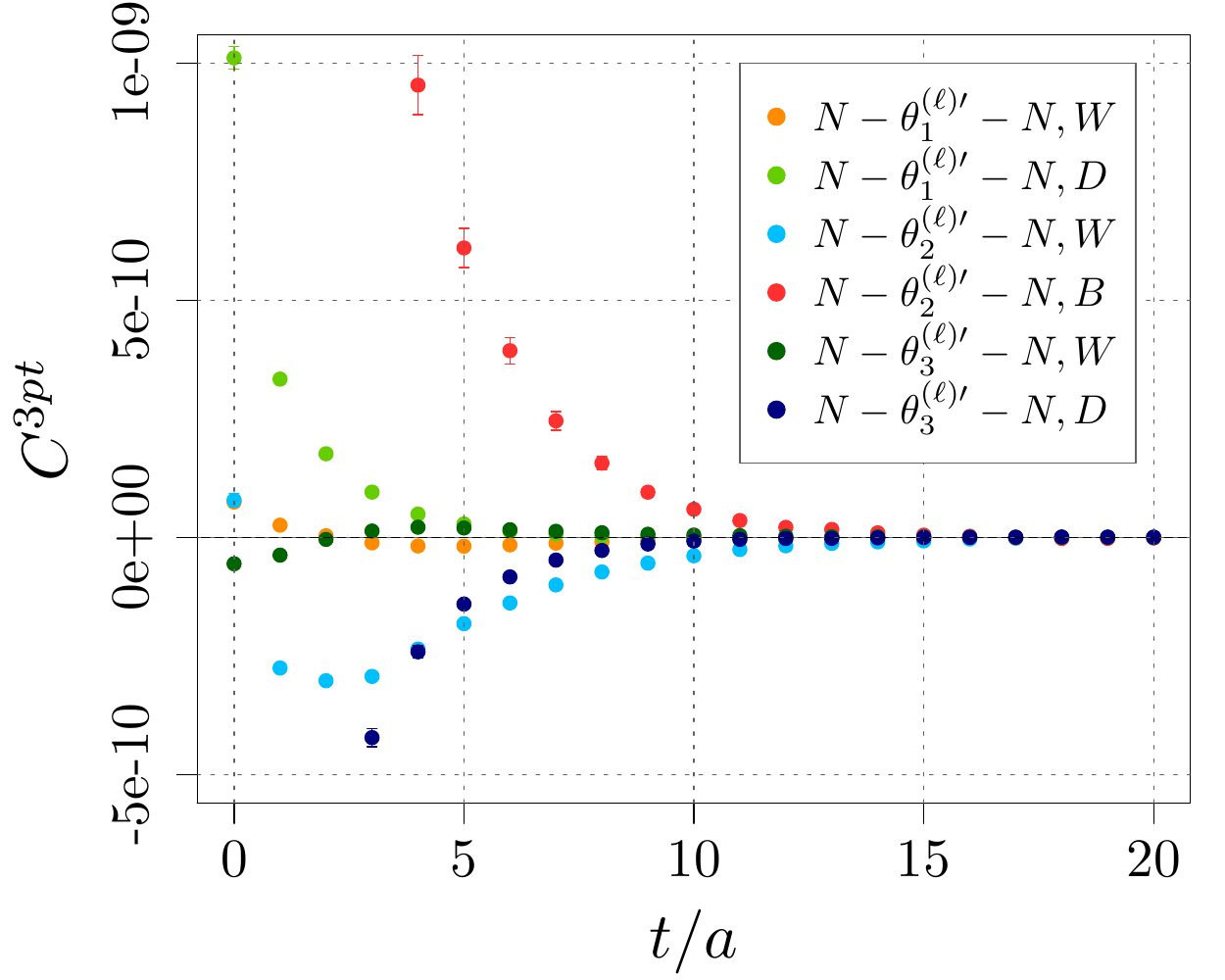}\label{fig:C3pt_light}}
\hfill
\subfloat[\bf{(b)}]{\includegraphics[width=0.497\textwidth]{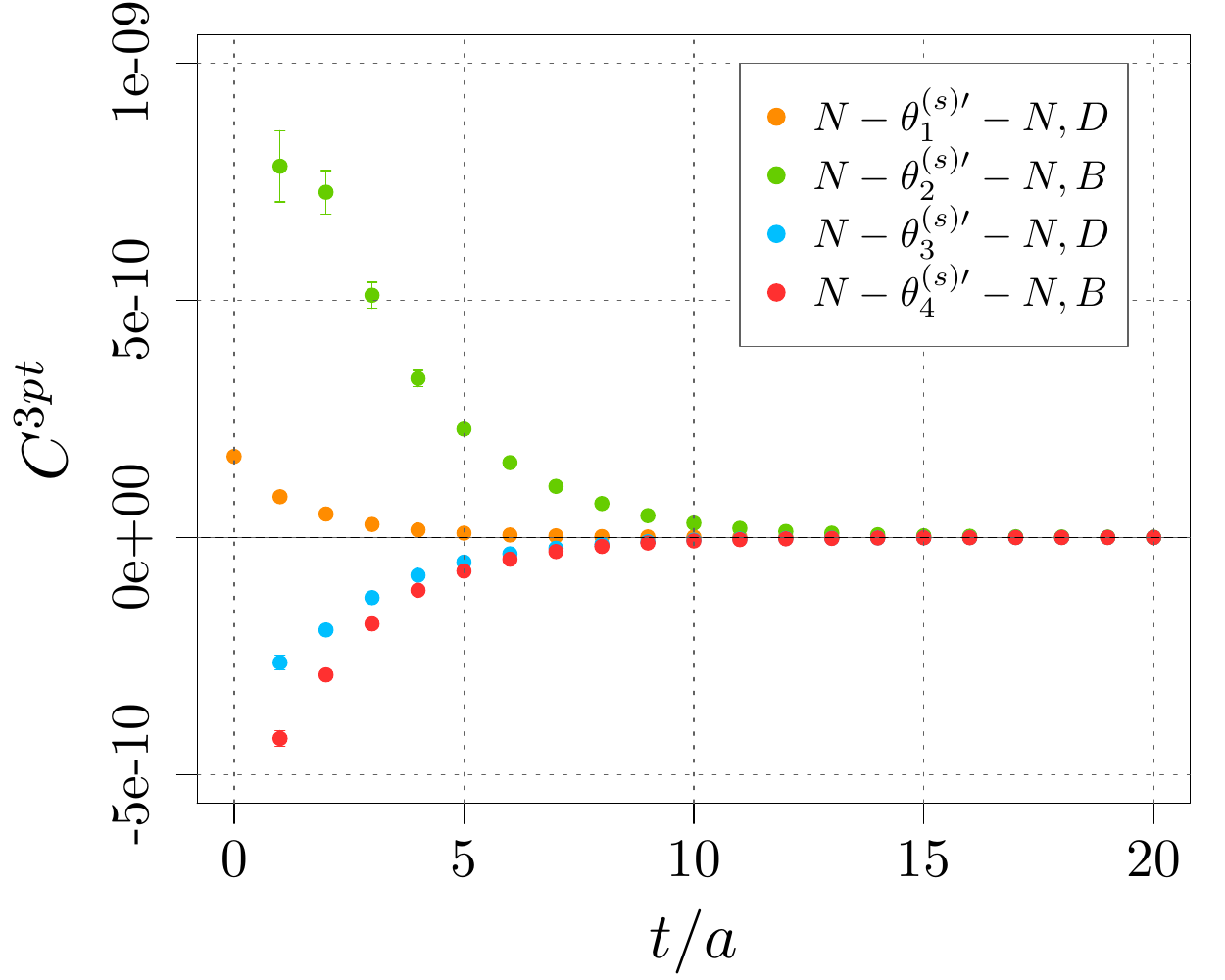}\label{fig:C3pt_strange}}
\hfill
\caption{Selected 3-point correlators with (a) light 4-quark operator $\theta_k^{(\ell)\prime}$ insertion and (b) strange 4-quark operator $\theta_k^{(s)\prime}$ insertion. The interpolator structure of the 3-point functions is specified in the legends.}
\label{fig:C3pt}
\end{figure}

A representative sample of the relevant 3-point correlators for the three different diagram types is shown in Fig.~\ref{fig:C3pt} for (a) the light quark sector and (b) the strange quark sector.
All correlators exhibit a clear signal which can be seen in Fig.~\ref{fig:C3pt} up to $t/a \lesssim 15$. 

\begin{figure}
	\centering
	\subfloat{
		\includegraphics[width=0.49\textwidth]{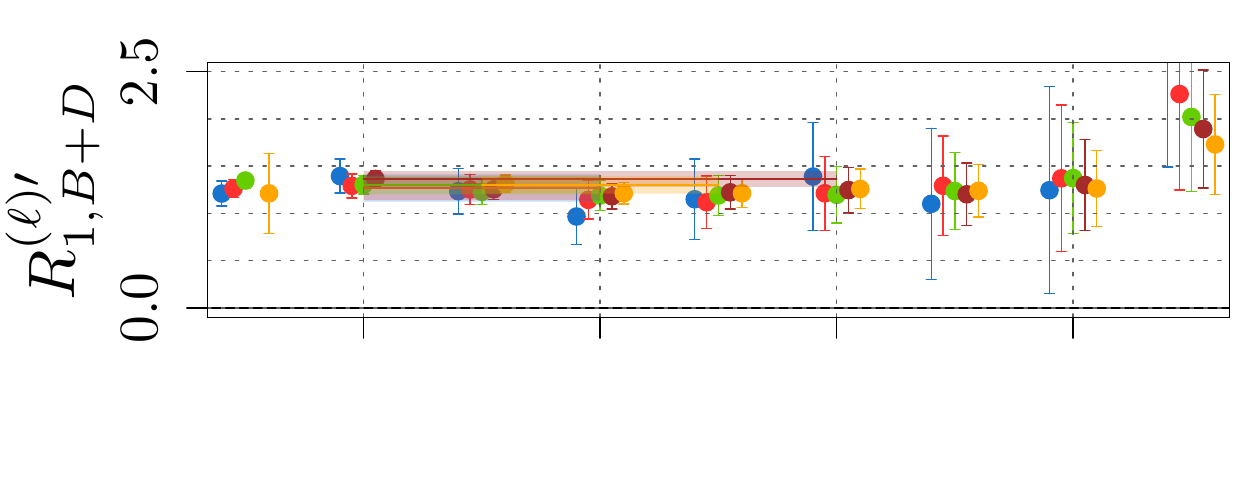}}
	\subfloat{
		\includegraphics[width=0.49\textwidth]{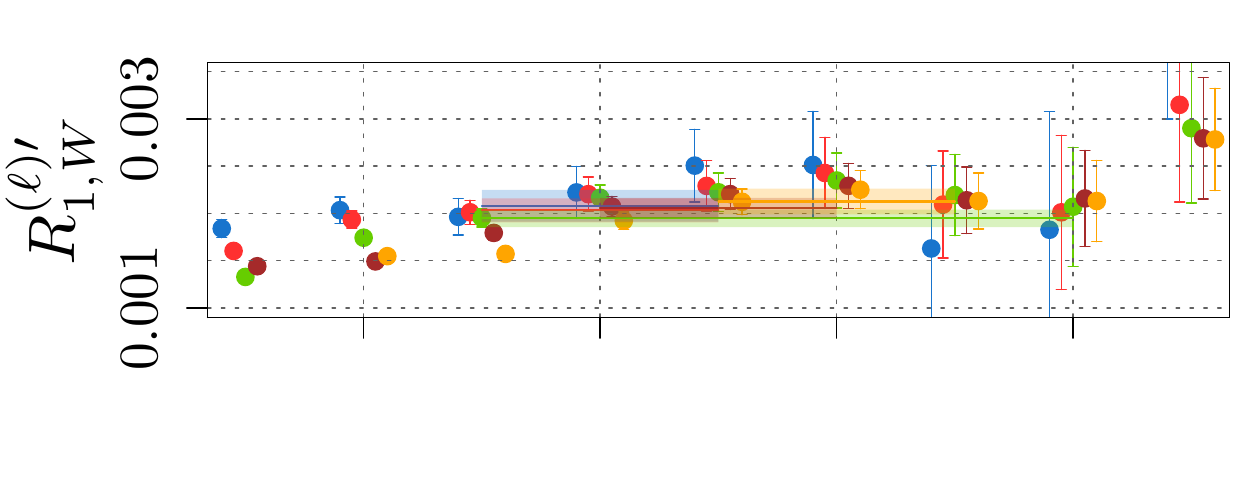}}\\
	\vspace{-0.9cm}
	\subfloat{
		\includegraphics[width=0.49\textwidth]{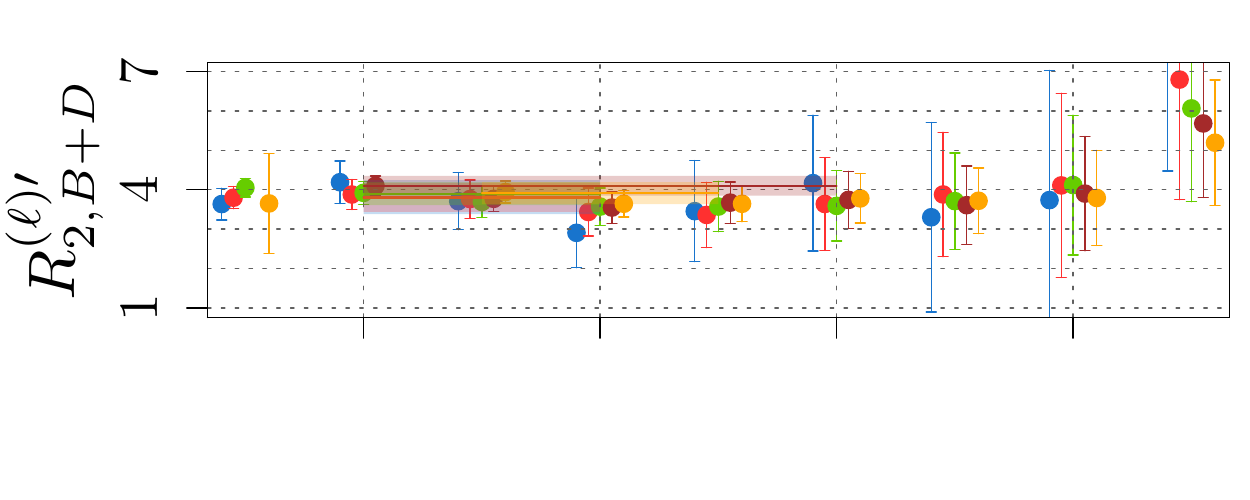}}
	\subfloat{
		\includegraphics[width=0.49\textwidth]{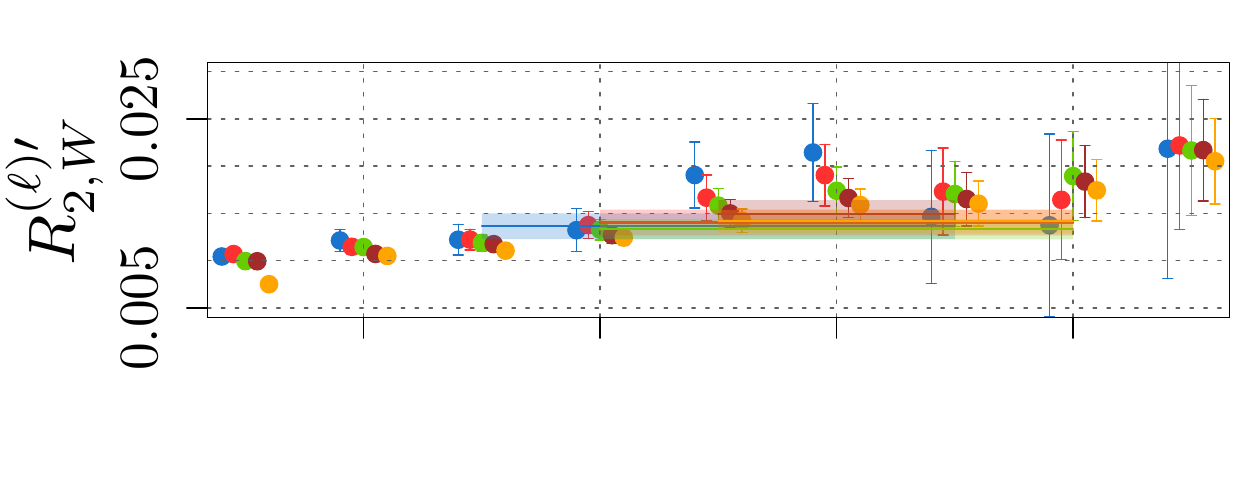}}\\
	\vspace{-0.9cm}
	\subfloat[\bf{(a)}]{
		\includegraphics[width=0.49\textwidth]{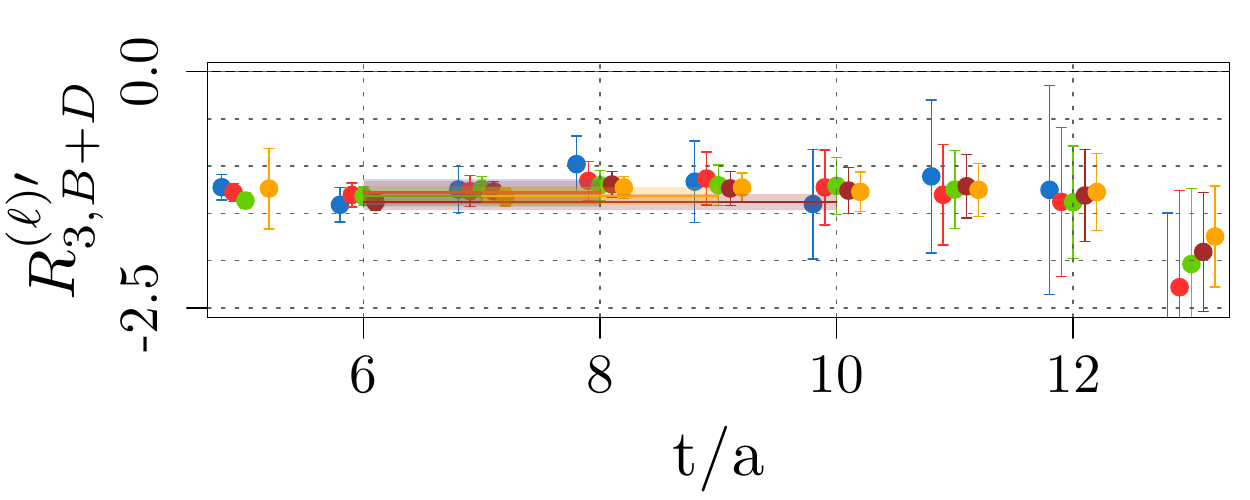}}
	\subfloat[\bf{(b)}]{
		\includegraphics[width=0.49\textwidth]{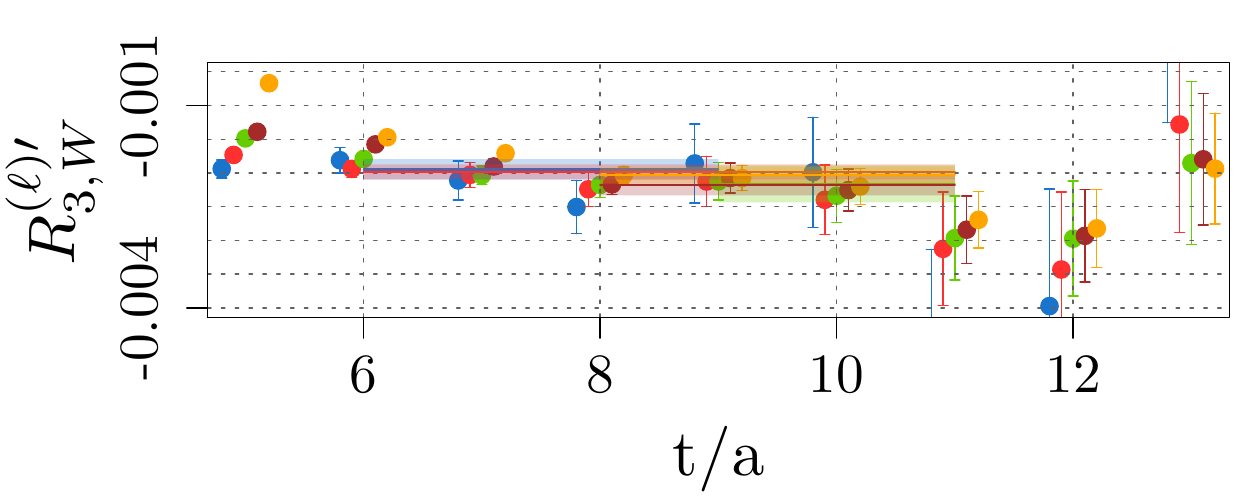}}
	\caption{Ratio plots for light 4-quark insertion. Here $R_{k, B+D}^{(\ell)\prime} = R_{k, B}^{(\ell)\prime} + R_{k, D}^{(\ell)\prime}$. Preliminary nucleon matrix elements are determined using constant fits in the plateau region. The different colors label different values of $\tau$, where the same color-coding applies as in Fig.~\ref{fig:h_pi1}.}
	\label{fig:ratios_light}
\end{figure}

\begin{figure}
	\centering
	\subfloat{
		\includegraphics[width=0.49\textwidth]{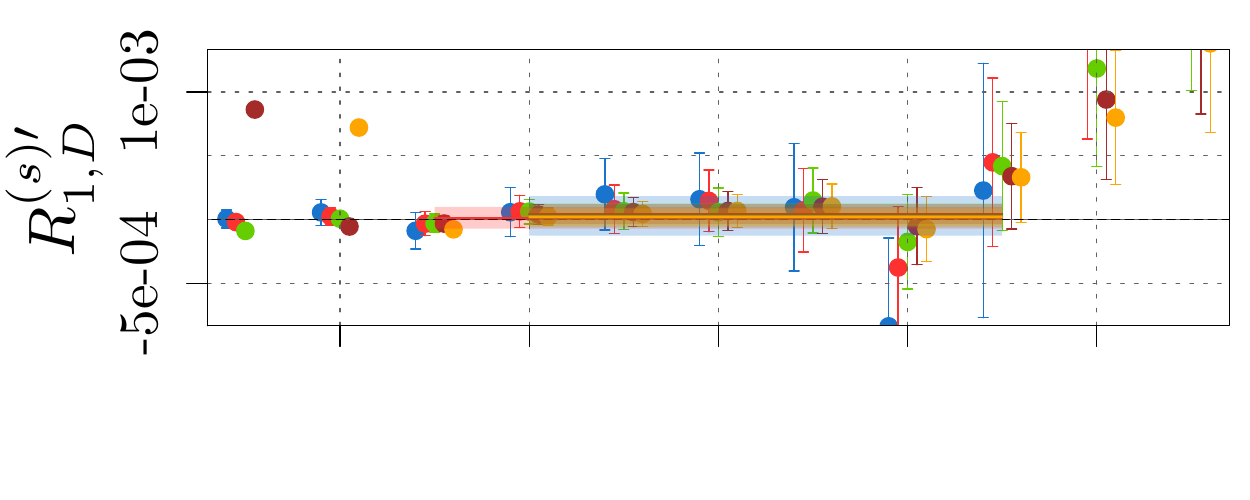}}
	\subfloat{
		\includegraphics[width=0.49\textwidth]{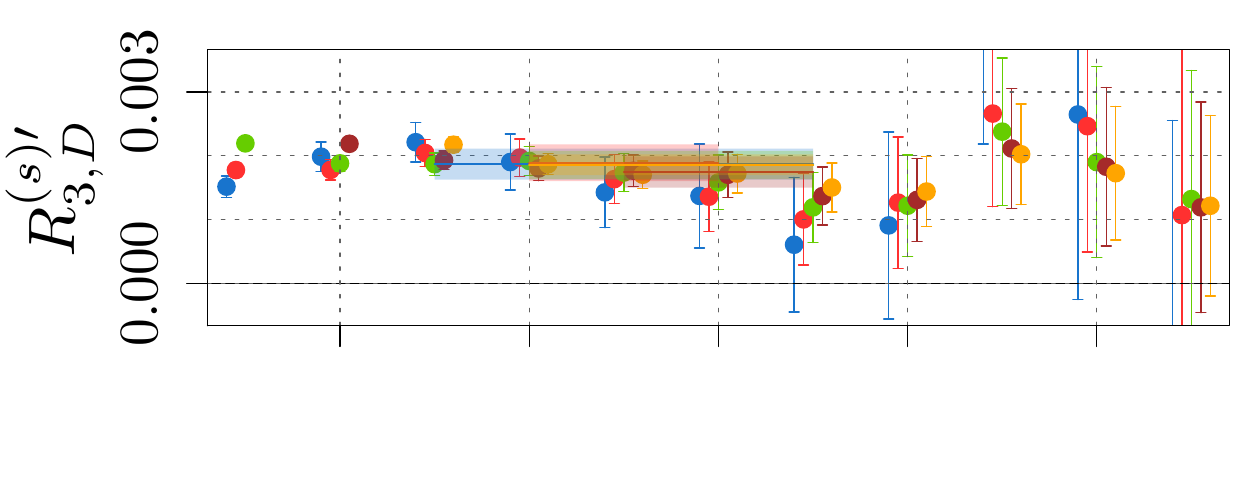}}\\
	\vspace{-0.8cm}
	\subfloat{
		\includegraphics[width=0.49\textwidth]{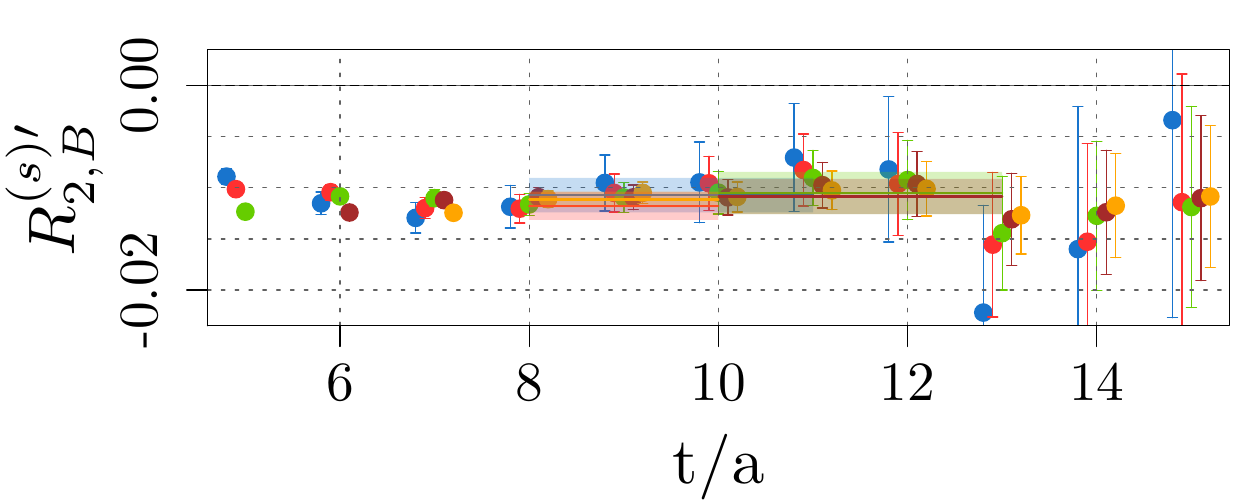}}
	\subfloat{
		\includegraphics[width=0.49\textwidth]{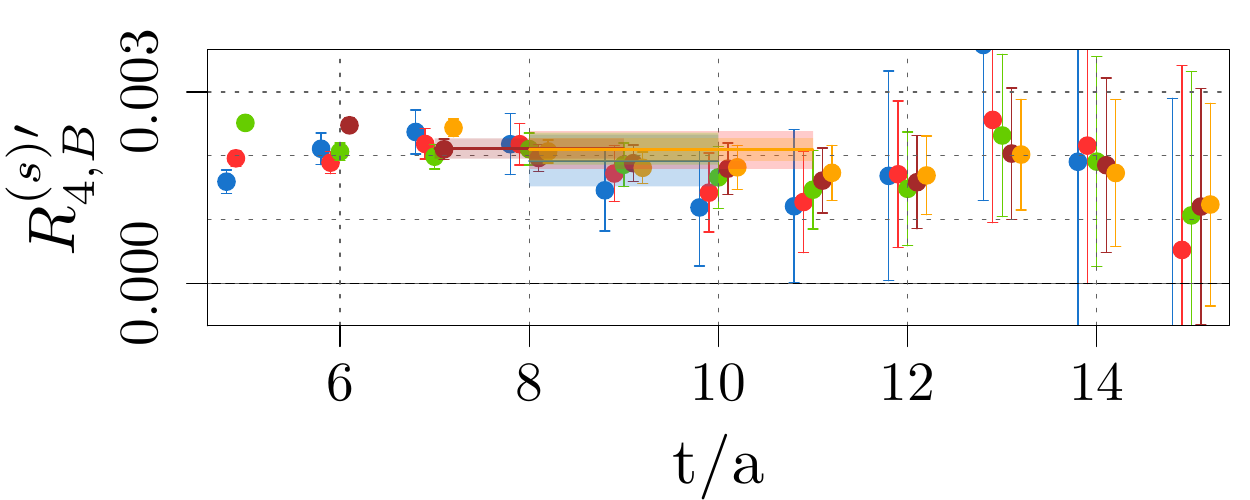}}
	\caption{Ratio plots for strange 4-quark insertion. Preliminary nucleon matrix elements are determined using constant fits in the plateau region. The different colors label different values of $\tau$, where the same color-coding applies as in Fig.~\ref{fig:h_pi1}.}
	\label{fig:ratios_strange}
\end{figure}

Preliminary results in the light and strange quark sectors for the contributions of the $B$- and $D$-type diagrams are summarized in Tab.~\ref{tab:matrix_elements_1}.
\begin{table}[htbp!]
	\centering
	\begin{tabular}{c|rrrr|rrrr}
		\hline\hline
		$\Bigg. k$ & $\tau$ & $\frac{\big.\sum_X \braket{N|\theta^{(\ell)\prime}_{k,X}|N}}{\big.2 m_N}$ & $\chi^2/\text{ndof}$ & $p$val & $\tau$ & $\frac{\big.\braket{N|\theta^{(s)\prime}_{k,X}|N}}{\big.2 m_N} / 10^{-2}$ & $\chi^2/\text{ndof}$ & $p$val\\
		\hline
		$\Big. 1$ & $5$ & $1.365(83)$ & $0.93$ & $0.44$ & $6$ & $0.002(7)$ & $0.95$ & $0.45$\\
		$\Big. 2$ & $5$ & $4.10(25)$ & $0.94$ & $0.44$ & $4$ & $-1.05(21)$ & $0.94$ & $0.42$\\
		$\Big. 3$ & $5$ & $-1.380(84)$ & $0.95$ & $0.43$ & $2$ & $0.187(24)$ & $0.93$ & $0.45$\\
		$\Big. 4$ & -- & -- & -- & -- & $3$ & $0.209(30)$ & $0.83$ & $0.48$\\
		\hline\hline
	\end{tabular}
	\caption{Preliminary nucleon matrix element results for the quark loop diagrams of type $B$ and $D$. For the light matrix elements labeled with $\ell$, $X=B, D$ holds. In the case of strange matrix elements labeled with $s$, for odd operator index $k$, $X=D$ and for even $k$, $X=B$. The light matrix element results are obtained from the plateau fits shown in Fig.~\ref{fig:ratios_light} (a) and the strange ones from the fits shown in Fig.~\ref{fig:ratios_strange}.}
	\label{tab:matrix_elements_1}
\end{table}
All these values are extracted from the ratio plots shown in
Figs.~\ref{fig:ratios_light} (a) and \ref{fig:ratios_strange} by
correlated constant fits. In the latter figures, the best fit lines for different $\tau$ values and the corresponding transparent error bands are drawn in color.
Note that the three different light operators have contributions of
both $B$- and $D$-type diagrams. Therefore, the combined matrix
elements are calculated, cf. left part of
Tab.~\ref{tab:matrix_elements_1}. The preliminary results for the
loop-free $W$-type diagram contributions are listed in
Tab.~\ref{tab:matrix_elements_2} and the corresponding best fit
results to the ratio data are shown in Fig.~\ref{fig:ratios_light}
(b).  

It is noteworthy that in the light quark sector the matrix elements
for $k=1$ and $k=3$ in the case of both the $B$- and $D$-type diagrams (Tab.~\ref{tab:matrix_elements_1}) and in the case of the $W$-type diagrams (Tab.~\ref{tab:matrix_elements_2}) match within errors except for the sign. Furthermore, in the strange quark sector, the matrix elements for $k=3$ and $k=4$ are indistinguishable within errors.

\subsection{Pion-Nucleon Coupling from $W$-Type Diagrams}

\begin{figure}[htbp!]
	\centering
	\includegraphics[width=0.62\textwidth]{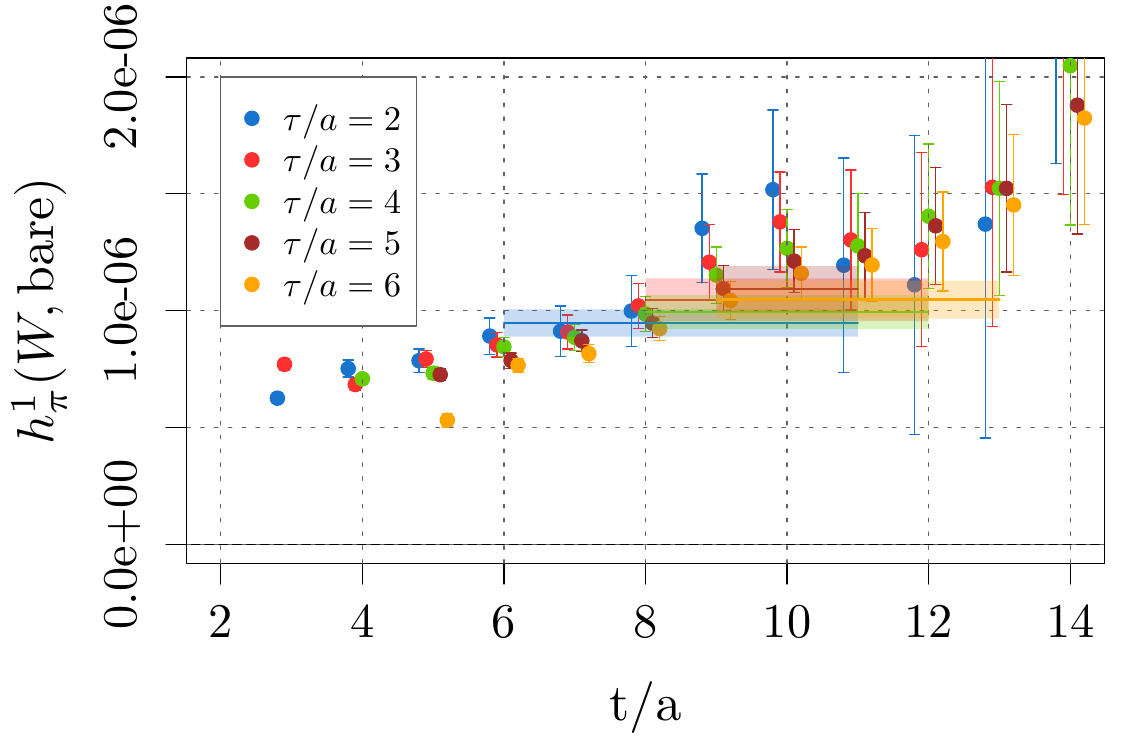}
	\caption{Preliminary results for the pion-nucleon coupling $h_\pi^1 (W, \text{bare})$ as a function of $t/a$ with variation of $\tau/a$. The pion-nucleon coupling results in Tab.~\ref{tab_h_pi1} correspond to the constant fit lines and error bands shown in this figure. All best fit lines lie in the plateau region with $6 \lesssim t/a \lesssim 13$.}
	\label{fig:h_pi1}
\end{figure}

Next, we determine the pion-nucleon coupling constant $h_\pi^1 (W, \text{bare})$ as a function of $t/a$ for different values of $\tau/a$, cf. Eqs.~(\ref{eq:FHT_ratio_2}) and (\ref{eq:h_pi1_only_W}). The result is shown in Fig.~\ref{fig:h_pi1}, along with the constant-fit lines in the plateau region and the corresponding error bands.
Here, we clearly see that the noise becomes so strong at larger $t/a$ values that the fit plateaus are formed only up to $t/a \gtrsim 13$. Therefore, we restrict the maximum fit range for the constant fits to $t/a \in [6, 13]$. The fit results associated with Fig.~\ref{fig:h_pi1} are listed in Tab.~\ref{tab_h_pi1}.
\begin{table}[htbp!]
\centering
\begin{minipage}[c]{0.49\textwidth}
\centering
	\begin{tabular}{crrrr}
		\hline\hline
		$\Bigg. k$ & $\tau$ & $\frac{\big.\braket{N|\theta^{(\ell)\prime}_{k,W}|N}}{\big.2 m_N}$ & $\chi^2/\text{ndof}$ & $p$val\\
		\hline
		$\Big. 1$ & $4$ & $1.95(9) \times 10^{-3}$ & $1.08$ & $0.37$\\
		$\Big. 2$ & $2$ & $1.36(14) \times 10^{-2}$ & $0.95$ & $0.44$\\
		$\Big. 3$ & $2$ & $-1.95(15) \times 10^{-3}$ & $0.93$ & $0.43$\\
		\hline\hline
	\end{tabular}
	\caption{Preliminary nucleon matrix element results for the loop-free $W$-type diagram obtained from the plateau fits shown in Fig.~\ref{fig:ratios_light} (b). Here, $k$ denotes the index of the operator at insertion.}
	\label{tab:matrix_elements_2}
\end{minipage}
\begin{minipage}[c]{0.49\textwidth}
	\centering
	\begin{tabular}{cccc}
		\hline\hline
		$\Big. \tau$ & $h_\pi^1(W, \mathrm{bare}) / 10^{-7}$ & $\chi^2/\text{ndof}$ & $p$val\\
		\hline
		$\Big.$2 & $9.47(59)$& $0.94$ & $0.46$\\
		$\Big.$3 & $10.46(92)$& $0.87$ & $0.48$\\
		$\Big.$4 & $9.94(73)$& $1.14$ & $0.34$\\
		$\Big.$5 & $10.93(98)$& $0.86$ & $0.42$\\
		$\Big.$6 & $10.48(81)$ & $0.83$ & $0.51$\\
		\hline\hline
	\end{tabular}
	\caption{Preliminary results for the pion-nucleon coupling obtained from the plateau fits shown in Fig.~\ref{fig:h_pi1}}
	\label{tab_h_pi1}
\end{minipage}
\end{table}
As expected from Eq.~(\ref{eq:FHT_ratio_2}), the fit results of
$h_\pi^1 (W, \text{bare})$ for the different $\tau/a$ values are
compatible within their errors. The best fit result with regard to both the
reduced $\chi^2$ value and the $p$ value is obtained with $\tau/a=2$.
In the following we use the result of this fit for the further discussion.

\section{Conclusion}

Our preliminary value for the pion-nucleon coupling $h_\pi^1 (W,
\text{bare})$ is given together with the experimental
result~\cite{Page:1987ak,NPDGamma:2018vhh}, and the LQCD result from Ref.~\cite{Wasem:2011tp}
in Tab~\ref{tab:h_pi1_results}. 
\begin{table}[htbp!]
\centering
\begin{tabular}{ccccc}
	\hline\hline
	$\Big.$ group & year & method & $M_\pi / \text{MeV}$ & $h_\pi^1 / 10^{-7}$\\
	\hline
	$\Big.$Page \textit{et al.} \cite{Page:1987ak} & 1987 & experiment & $140$ & $0.4_{-0.4}^{+1.4}$\\
	$\Big.$Wasem \cite{Wasem:2011tp} & 2012 & LQCD, $\mathcal{L}_{PV}$ & $390$ & $1.10(51)$\\
	$\Big.$NPDGamma \cite{NPDGamma:2018vhh} & 2018 & experiment & $140$ & $2.6(1.2)$\\
	$\Big.$ our work & 2022 & LQCD, $\mathcal{L}_{PC}$& $260$ & $9.47(59)$\\
	\hline\hline
\end{tabular}
\caption{Selected experimental and numerical results for $h_\pi^1$}
\label{tab:h_pi1_results}
\end{table}
For this preliminary determination of $h_{\pi}^1$, we find a result of the same order of magnitude as that of the NPDGamma collaboration~\cite{NPDGamma:2018vhh} and as the result from Ref.~\cite{Wasem:2011tp}. However, one has to keep the
limitations of our calculation (no quark loop diagrams, no renormalization, single lattice
spacing and pion mass value) in mind.

In addition, we have extracted for the first time also the light and
strange quark-loop diagrams, see Tab.~\ref{tab:matrix_elements_1},
with $\lesssim 10\%$ relative error. We think that this precision in
the bare matrix elements is an advantage of the method applied and
tested here.

On the one hand we are currently investigating the pion mass dependence
of the quantities of interest. 
On the other hand, in our lattice calculation we expect mixing with
lower dimensional operators, whose investigation is underway based on
the gradient flow method.

\section*{Acknowledgments}

We gratefully acknowledge the generous support by the Deutsche Forschungsgemeinschaft (DFG, German Research Foundation) and the NSFC through the funds provided to the Sino-German Collaborative Research Center CRC 110 "Symmetries and the Emergence of Structure in QCD" (DFG Project-ID 196253076 - TRR 110, NSFC Grant No. 12070131001).

\bibliographystyle{h-physrev5} 

\end{document}